\documentstyle[aps,pre]{revtex}          
\tighten                                  
\begin{document}                          
\draft                                    
\twocolumn                               

\title{Molecular ordering of precursor films during
spreading of tiny liquid droplets}

\author{M. Haataja$^{1,2}$,
J. A. Nieminen$^{1}$, and T. Ala--Nissila$^{1-3}$}

\address{
$^1$Department of Physics, Tampere University of Technology,\\
    P.O. Box 692, FIN--33101 Tampere, Finland}

\address{
$^2$Research Institute for Theoretical Physics,
\\P.O. Box 9 (Siltavuorenpenger 20 C),
FIN--00014 University of Helsinki, Finland
}

\address{
$^3$Department of Physics, Brown University, Box 1843,
Providence, R.I. 02912, U.S.A.
}

\date{21.7.1995}

\maketitle
\narrowtext

\begin{abstract}

In this work we address a novel feature
of spreading dynamics of tiny liquid droplets on
solid surfaces, namely the case where
the ends of the molecules feel different interactions to the
surface. We consider a simple model
of dimers and short chain--like molecules which cannot
form chemical bonds with the surface.
We study the spreading dynamics by Molecular Dynamics techniques.
In particular, we examine the microscopic
structure of the time--dependent precursor film and find that in
some cases it can exhibit a high degree of local order.
This order persists even for flexible chains.
Our results suggest the possibility of extracting
information about molecular interactions from the structure of the
precursor film.

\end{abstract}


Since the discovery of dynamical
layering by Heslot {\it {et al.}} \cite{Hes89a},
there has been increasing interest in phenomena
occuring at microscopic length--scales during
spreading. The beautiful experiments of
Refs. \cite{Hes89a,Hes89b,Hes90,Alb92}
reveal that both the
molecular structure of the liquid and the type of
substrate used influence density
profiles of the droplets. For example, thickness profiles
of tetrakis (2--ethylexoxy)--silane and polydimethylsiloxane
(PDMS) droplets on a silicon wafer exhibit strikingly different
shapes under spreading \cite{Hes89a}. Tetrakis exhibits clearly
observable dynamical layering, while the spreading of PDMS
proceeds by a fast evoluting precursor layer of one molecular
thickness. Furthermore, the experiments of Heslot {\it {et al.}}
\cite{Hes89b} and Valignat {\it {et al.}}
\cite{Val93} also demonstrate the important role of
liquid--surface interactions
in determining the morphology of spreading droplets in microscopic
scales.

Because of these experiments theoretical work has gained new interest.
Analytic theories to date deal with dynamical
layering only \cite{Abr90,deG90}.
A number of computer simulations have been performed
\cite{Hei91,Yan91,Nie92,DeC93,Nie94,Ven94,Wag95,DeC95}
to study the dynamics of spreading with some emphasis on the
molecular structure of the liquid.
In particular, using Molecular Dynamics (MD) simulations
it was concluded in Refs. \cite{Nie94} that
the chain--like nature of the molecules
can influence the structure of the precursor layer.

The molecular structure of
thin layers is of practical importance, too.
One way of controlling
the surface energetics of a solid
substrate is to use grafted molecules that adsorb on it,
sometimes forming chemical bonds and
brushed layers \cite{Mil91}. Such surfaces have
applications in e.g. coating and lubrication.
Another important class of systems comprises amphiphilic molecules
such as detergents where a strong asymmetry of interactions
causes layered structures to form \cite{Isr92}.
Spreading dynamics of such molecules is then of
particular interest in trying to understand how these
layers form, and how well ordered they are.

Motivated by these considerations,
we have
chosen to study the dynamics of spreading of simple
models of short, asymmetrically interacting
molecules using MD simulations. This method is convenient
since microscopic information about the structure
and properties of the spreading layers is readily
available in real time.
Our model follows Ref. \cite{Nie94} where
short chain--like molecules
interact with each other via the Lennard--Jones (LJ) potential
$V(r)= 4 \epsilon_{f} ((\sigma_f/r)^{12}-(\sigma_f/r)^{6})$,
with parameters \(\sigma_{f}\) and
\(\epsilon_{f}\) for the width and depth of the potential,
respectively. The \(n\)--mers comprise \(n\) LJ particles
interconnected by a very rigid pair potential
\(V_{c} = (1/2)k(r-r_{0})^{2}\),
where \(k = 10000 \epsilon_{f} / \sigma_{f}^{2}\).

The surface is modelled by a
flat, continuum LJ material with a pairwise LJ interaction
integrated over the volume $z \leq 0$ to
obtain an interaction
potential $V_i(z) = -A/z^3 + B/z^9$, where
$A=(2 \pi/3) \epsilon_s \sigma_s^6 \rho$, and
$B=(4 \pi/45) \epsilon_s \sigma_s^{12} \rho$ \cite{Nie94},
with $\rho \sim 10^{30} \ m^{-3}$.
Table I shows the parameters.
Interaction asymmetry comes about through the choice of
the potentials: \(V_I\) for one end of the chain
(called the grafted end) and either \(V_{II}\) or
$V_{III}$ for the other. The potential
$V_{I}$ is much steeper and deeper than $V_{II}$
and $V_{III}$, both of which are of equal depth.
In addition, the equilibrium distance of the surface
potential minimum as
seen by the monomers is the same for $V_{I}$
and $V_{II}$ (the {\it ordinary case}),
but for $V_{III}$ (the {\it shifted case})
the minimum distance is about two bond lengths
further away from the surface. For the shifted
case this means that individual
dimers tend to lie
{\it perpendicular} to the surface.

The simulations are done at
a constant temperature of \(kT_{s}=0.8 \epsilon_{f}\),
which is well above the triple point of an LJ material \cite{Nie94}.
The dynamics is described by the
equations of motion with a Nos\'e--Hoover
thermostat \cite{Nie94,No84,Hu85}.
They are solved using a modified
velocity Verlet algorithm \cite{AlTi}.
The Hamaker constant $A_H$ is determined by the
effective bond length $b_{l}$ of the dimers; {\it e.g.} for
$b_{l} \approx 100$ {\AA},
$A_H \approx 10^{-18} J$ which is a realistic value for surfaces
\cite{Isr92}. Assuming that the effective mass
of the molecules is about $10^5 \ amu$ \cite{Val93},
the time step in reduced units (r.u.) is
$t_r \approx 10^{-13}$ seconds.

Fig. 1 shows snapshots of typical evolution of dimer
droplets during spreading for the two cases.
The random initial state of the ridge--shaped
droplet is first carefully
prepared, after which the surface interaction is turned on
and spreading occurs perpendicular to the axis of the ridge.
The dimers above the surface are
mainly oriented perpendicular to it due to
the stronger attraction of the grafted end.
Dimers on the surface, however,
behave quite differently. In the ordinary
case the precursor film appears rather
rough except very close to the edges of the droplet
where the molecules have enough room to lie flat.
In the shifted case the situation changes dramatically
and the precursor film appears very well ordered at all stages
of spreading.

To quantify these observations, we have followed the time
evolution of various density--density autocorrelation
functions \cite{AlTi}. In particular, we have defined the
{\it precursor pair correlation function} as follows:
\begin{equation}
g(| \vec {r_i} - \vec {r_j}|,t) = N(t)^{-2} <\sum_{i} \sum_{j \neq i}
\delta(\vec {r_{i}})
\delta(\vec {r_{j}}) >\,\,\mbox{.}
\end{equation}
The total number of molecules within the precursor
film is denoted by $N(t)$,
$\vec {r_i}(t)$ is the
position vector of the centre--of--mass of
the $i^{th}$ dimer, and the summations go over the
dimers in the precursor layer. This function has been
computed during the spreading dynamics.
After a short inital transient, it assumes
a characteristic steady--state shape for the times studied
here. For the ordinary case, $g$
has fairly sharp peaks corresponding to
nearest neighbour dimers only
which indicates that the layer is
disordered and liquid--like. In the shifted case, however,
clear and sharp peaks can be observed corresponding
up to about fourth or
fifth nearest neigbour dimers. The precursor layer
in this case has a high degree of local order
even at these elevated temperatures.
We have also followed the evolution of an orientational
distribution function which provides us with information
about the relative orientations of neighbouring dimers.
For the shifted case a sharp peak is observed corresponding to
parallel orientations, which indicates that the orientations of
neighbouring dimers are strongly correlated, too.
In both cases the
very late time regime corresponds to a diffusively
thinning monolayer.

Fig. 2 shows typical smoothed density profiles of dimers
at three different times. In the ordinary
case the profiles are fairly smooth and rounded,
and no layering is present. At later times a step develops
at the edges of the film where dimers tend to lie flat
on the surface. In the shifted case the edge of the
precursor film always remains very sharp and well--defined.
It is conceivable that tendency towards layering on top
of the precursor layer might be
observed for larger droplets in this case.

We have also studied the time--dependence of the
precursor film width $w(t)$.
Fig. 3 shows a comparison between the two cases.
They are very similar, with a more rapid
``almost linear'' ($\sim t^{0.9}$) region
followed by a crossover
to slower ``diffusive'' ($\sim t^{0.5}$)
behaviour. This is characteristic of the truly
microscopic droplets studied here and has been observed
in Refs. \cite{Nie92,Nie94} as well. For our cylindrical geometry,
within the first region $w(t) \propto N(t) \propto t^{0.9}$,
which for a true $3-d$ droplet
means that its radius $R(t) \sim \sqrt{N} \sim t^{0.45}$
since the flux into the first layer remains roughly constant.
This overall $R(t) \sim t^{0.5}$ behaviour
is in agreement with experiments \cite{Hes89b,Hes90,Alb92,Val93}.

Furthermore, it is possible to estimate the
corresponding effective diffusion coefficients in the
two regimes.
In the ordinary case for the ``truly'' diffusive late--time
regime we find $D_{\ell} \sim 10^{-6} \
m^2/s$ \cite{diff}.
For the shifted case $D_{\ell}$
is about an order of magnitude smaller
because the activation energy arising from neighbouring dimers
is much larger due to
the close--packed geometry of the layer.
Most importantly, however, estimating a ``diffusion''
coefficient $D_e$ from the early--time
(almost) linear region, we
find that  $D_{e}/D_{\ell} \approx 100$.
This is in surprisingly good quantitative agreement with
recent experiments, where the same ratio was measured
between early--time and late--time ``diffusion'' coefficients
\cite{Val93}.

Additional studies of flexible chains consisting
of four and eight particles where one of the ends is grafted
with $V_I$ reveal that
the spreading becomes similar to
the results of Refs. \cite{Nie94}
without any asymmetry of interaction, provided that the equilibrium
orientation of an individual chain is parallel to the surface. In the
case of a shifted potential, however, in which the equilibrium orientation
of an individual chain is perpendicular to the surface, compact
and well--ordered structure of the precursor
layer still persists even for completely flexible chains.
Complete results with different
shifted potentials will be published elsewhere.

Due to the microscopic nature of our droplets, it is difficult
to make comparisons with experiments.
A spreading experiment
with grafted polymer chains has been performed
by Valignat {\it {et al.}} \cite{Val93}. In some cases,
the molecules become anchored on the surface and spreading
stops. However, some of their density profiles do qualitatively
resemble ours. For example, in the case of a microdroplet of PDMS--OH
on a silicon wafer covered with a layer of behenic acid,
a well--defined and sharp precursor film develops similar to our
shifted case. We are not aware of any experiments probing
microscopic order within the precursor film.
However, it is interesting to note that in the
experiments of Ref. \cite{Isr88}, {\it equilibrium} ordering of
thin layers of a simple liquid between two corrugated
surfaces was observed.

In conclusion, we have shown further evidence
of how the surface--chain interactions
can play a significant role in determining the microscopic
structure of the precursor layer.
In particular, through a simple model
we have demonstrated how the internal structure of
the film can depend on the asymmetry of surface interactions
between the chain ends. This can lead to a high degree of
local order if the molecules prefer a vertical equilibrium
orientation. Additionally, our model reproduces the
experimentally observed more rapid spreading at the
early stages, which crosses over to slower diffusion at late times.
Based on our results we suggest the possibility of
obtaining information about the nature of
these interactions by monitoring the density
profile and precursor film during spreading.

\pagebreak
\cleardoublepage
\pagebreak

\begin{table}[t]
  \vspace{0.5cm}
  \begin{tabular}{||c|c|c||}
  $\epsilon_{s}$      &   $\sigma_{s}$   &    Potential
     \\ \hline
  1.0 $\epsilon_{f}$  & 5.0 $\sigma_{f}$ &    $V_{I}$
      \\ \hline
  0.06 $\epsilon_{f}$ & 5.0 $\sigma_{f}$ &    $V_{II}$
      \\ \hline
  0.02 $\epsilon_{f}$ & 7.3 $\sigma_{f}$ &    $V_{III}$
      \\
  \end{tabular}
  \vspace{0.5cm}
  \caption{Surface--monomer interaction parameters for the two
  cases discussed in the text.
  }
  \vspace{0.5cm}
\end{table}

\cleardoublepage
\pagebreak
\begin{center}
\Large
{\sc figure captions}
\end{center}

\normalsize

\vspace{2.0cm}

Fig. 1. Snapshots of a dimer droplet for the ordinary and the shifted cases.
        The grafted end is represented by a filled circle.
 (a) Initial configuration for both cases (as seen along the axis of the
ridge).
(b) The ordinary case at $ t = 30000\, r.u.$,
and (c) the shifted case at $ t = 30000\, r.u.\,$.
Differences in the morphology of the
precursor layer are clearly visible.

\vspace{1.0cm}

Fig. 2. Smoothed density profiles for the two cases: (a) the
ordinary case, and (b) the shifted case.
The shoulders at the edge of the precursor film in (a)
are due to dimers that fall flat on the surface.

\vspace{1.0cm}

Fig. 3. Width of the precursor film $w(t)$ vs. time for the
two cases discussed in the text. This data is for 1525 dimers.
Note the considerably slower late--time behaviour of the shifted
case.


\begin{thebibliography}{99}
\bibitem{diff}
This is somewhat larger than the measured values
(see e.g. Refs. \cite{Hes89b,Hes90,Val93})
which is not surprising since in our units
$T \approx 1600 \ K$, and there are no surface
diffusion barriers.
\end{thebibliography}
\end{document}